\documentclass[manuscript]{aastex}
\usepackage{natbib}
\usepackage{psfig}
\usepackage{wasysym}
\usepackage{txfonts}
\def\AV{\mbox{A$_{\rm V}$}}

\def\nH2{\mbox{${\rm n}(\HH$)}}
\def\enH2{\mbox{$n_{(\HH$)}}}

\def\pccc{~{\rm cm}^{-3}} 
\def\ccc{~{\rm cm}^3} 
\def\pcc {~{\rm cm}^{-2}}
\def\Tstar {\mbox{${\rm T}_{\rm R}^*$}}
\def\Tsub#1 {\mbox{${\rm T}_{\rm #1}$}}
\def\TK  {\Tsub K }

\def\arcsec{\mbox{$^{\prime\prime}$}} \def\arcmin{\mbox{$^{\prime}$}}

\def\degr{$^{\rm o}$}
\def\p{\mbox{$^+$}}
\def\z{\mbox{${^0}$}}

\def\TWC{\mbox{{$^{12}$C}}} 
\def\THC{\mbox{{$^{13}$C}}} 
\def\FTN{\mbox{{$^{15}$N}}} 
\def\TFS{\mbox{{$^{34}$S}}}

\def\h13cop{\mbox{{H$^{13}$CO\p}}}

\def\c3h2{\mbox{C$_3$H$_2$}}
 
 \def\R0{R$_0$}
\def\G0{\mbox{G$_0$}}

\def\ddeg{{}^\circ\kern-.1em}

\def\kms{\mbox{km\,s$^{-1}$}}
\def\ps{\mbox{s$^{-1}$}}

\def\E#1 {$10^{#1}$}
\def\E#1 {E{#1}}
\def\P#1,{$\nH2\TK~=~#1\times~10^4\pccc$~K}
\def\ec#1,#2,#3,{#1\,(#2)\E{#3}}

\def\H3{\mbox{H$_3$}}

\def\ammon{\mbox{N\H3} }

\def\zetaMolH{\mbox{$\zeta_\HH$}}

\def\RH2{\mbox{R$_{\rm G}$}}
\def\g13{\mbox{g$_{13}$}} 

\def\RC{\mbox{${\rm R}_{12/13}$}}

\def\kHeH2{\mbox{$k_{ He-\HH}$}} 
\def\kHeCO{\mbox{$k_{He-CO}$}} 
\def\kjCp{\mbox{$k_{j-C\p}$}} 
\def\kf{\mbox{$k_{f}$}} 
\def\kr{\mbox{$k_{r}$}} 
\def\tim#1,#2{\mbox{{$#1\times10^{#2}$}}}
\def\ft{\mbox{$\sqrt{(300/\TK)}$}}

\newcommand{\emm}[1]{\ensuremath{#1}}   
\newcommand{\emr}[1]{\emm{\mathrm{#1}}} 

\newcommand{\hcop}{\emr{HCO^+}} 
 
\newcommand{\HH}{\emr{H_2}}
\newcommand{\cotw}{\emr{^{12}CO}}
\renewcommand{\coth}{\emr{^{13}CO}}
\newcommand{\coei}{\emr{C^{18}O}}

\shorttitle{Where has all the \THC \p gone?}
\shortauthors{Liszt \& Ziurys}

\begin{document}


\title{Carbon isotope fractionation and depletion in 
TMC1\thanks{Based on observations obtained with the ARO Kitt Peak 12m
 telescope.}}


\author{H. S. Liszt}
\affil{National Radio Astronomy Observatory \\
            520 Edgemont Road,
           Charlottesville, VA,
           22903-2475}
\and

\author{L. M. Ziurys}
\affil{Arizona Radio Observatory \\ 
           University of Arizona, 
           933 N. Cherry Avenue, 
           Tucson, AZ 85721}

\email{hliszt@nrao.edu}



\begin{abstract}
\TWC/\THC\ isotopologue abundance anomalies have long been 
predicted for gas-phase chemistry in molecules other than CO and  
have recently been observed in the Taurus molecular cloud 
in several species hosting more than one carbon atom, i.e. CCH, CCS, CCCS
and HC$_3$N.  Here we work to ascertain whether these isotopologic 
anomalies actually result from the predicted depletion of the \THC\p ion
in an oxygen-rich optically-shielded dense gas, or from some other more 
particular mechanism or mechanisms.  We observed $\lambda$3mm emission from 
carbon, sulfur and nitrogen-bearing 
 isotopologues of HNC, CS and \HH CS at three positions in Taurus 
 (TMC1, L1527 and the \ammon peak) using the ARO 12m telescope.
We saw no evidence of \TWC/\THC\ anomalies in our observations.
Although the pool of C\p\ is likely to be depleted in \THC, 
\THC\ is not depleted in the general pool of carbon outside CO, which probably
exists mostly in the form of C\z.  The observed isotopologic abundance 
anomalies are peculiar to those species in which they are found.
\end{abstract}



\section{Introduction}

Observations of isotopically-substituted versions of interstellar 
molecules -- their so-called isotopologues -- have become
the usual way of deducing atomic isotopic abundance ratios for 
common products of stellar nucleosynthesis -- C, N, O, S, Si, Cl etc 
in the interstellar medium (ISM), as summarized by \cite{Wan80}
 and \cite{Wil99}.  However, the 
relative abundances of particular isotopologues can differ from those 
inherent in the gas at large owing to peculiarities of their chemistry
\citep{WatAni+76,Wat77,LanGra+84}, and recognizing or explaining anomalous 
isotopologic abundance ratios is an ongoing challenge for astrochemistry.

The most commonly-observed example of this phenomenon is with CO in diffuse
clouds, where the the CO/\coth\ ratio may be larger or smaller than the 
inherent C/\THC\ ratio in the ambient gas depending on whether selective
photodissociation \citep{BalLan82} or chemical fractionation \citep{WatAni+76}) 
dominates
\citep{WilMau+92,LisLuc98, SonWel+07,BurFra+07,SheRog+08,Lis07,
VisVan+09}.  Selective photodissociation (increasing CO/\coth)
occurs in gas seen toward  a few hot stars while chemical fractionation 
(behaving oppositely) is a more general phenomenon that is enhanced at somewhat 
higher CO column densities.  Chemical fractionation is also now seen in the 
envelopes of dark clouds \citep{GolHey+08}.  

Gas-phase molecular cloud chemistry actually makes a rather general prediction 
about isotopologic abundances in strongly shielded regions: CO and molecules 
that form directly from it (\hcop, \HH CO and CH$_3$OH) should show a common 
C/\THC\ ratio that reflects the composition of the gas reservoir, while other 
species that form from the pool of free carbon outside CO should be strongly 
depleted in \THC\ \citep{Wat77,Lis78,LanGra+84}.  This  occurs because CO 
is the repository of the overwhelming majority of the 
gas-phase carbon\footnote{at least as long as [O] $>$ [C]} and carbon is 
liberated from CO mainly as the dissociation product of destruction of CO 
by a small quantity of cosmic-ray ionized He\p\ 
(i.e. He\p + CO $\rightarrow$ C\p\ + O + He).  The resulting
C\p\ ions quickly interact with the ambient CO, and the fact
that \coth\ is more strongly bound than \cotw\ results in a preferential
deposition of \THC\ into \coth\ through the reaction 
\THC\p\ + \cotw\ $\rightarrow$ \TWC\p\ + \coth\ + 34.8 K 
(\cite{WatAni+76}, see Appendix A here).  
In this way \THC\p\ preferentially disappears from the pool of C\p\ and 
from the general pool of carbon available to form most species.  The 
C\p/\THC\p\ ratio increases by a factor approaching 
exp(34.8/\TK) but the CO/\coth\ ratio changes little 
because the C\p/CO abundance ratio is so small, below $10^{-3}$.

It quickly became apparent that there was some tension between this 
prediction and observations of interstellar gas showing C/\THC\ 
ratios that were, if anything, {\it below} the terrestrial value of 
C/\THC\ = 89.  Now it is acknowledged that the local interstellar carbon 
isotope ratio is in the range 60-70, but at the time, \cite{Wat77} 
suggested a mechanism involving selective depletion of carbon onto 
grains, offering the possibility of lowering the C/\THC\ ratio.  Detailed 
time-dependent models showed very complex behaviour in the isotopologic 
ratios but largely failed to bear out this suggestion \citep{Lis78}, 
finding that \THC\ enhancement outside CO  only occured as molecules 
disappeared from the gas.

Because of the possibility of fractionation in fully molecular gas,
some observers changed the focus of their work to concentrate on 
measurement of carbon isotope ratios using only the isotopologues of 
carbon monoxide \citep{LanPen93} but it seems fair to say that anomalously 
large C/\THC\ ratios have never been seen in the molecular 
cloud-H II region complexes that are typically used for isotope 
determinations \citep{Wan80,Wil99}.  Indeed, a recent survey of 
the carbon isotope ratio in CN does not seem to have been affected 
\citep{MilSav+05} and \cite{TerCer+10} report C/\THC\ = 45$\pm$ 20, 
S/\TFS = 20 $\pm$ 6 from a survey of sulfur-bearing carbon chain 
molecules in Ori KL.  Fractionation effects may be  minimized 
at the somewhat higher temperatures of clouds near H II regions but
are not entirely eliminated.  The absence of observable effects
from the predicted carbon fractionation chemistry has been a lingering, 
if not entirely obvious, mystery. 

By contrast, anomalies of two kinds have now been noted in cold gas 
in/near the cyanopolyyne peak in the Taurus Molecular Cloud (TMC) for 
species containing several carbon atoms, and \cite{SakIke+07} and 
\cite{SakSar+10} appealed to the work of \cite{LanGra+84} for an explanation.
$\THC$ is strongly but very unequally lacking in both of the 
singly-substituted \THC\ isotopologues of CCH \citep{SakSar+10}
and is deficient in \THC CS but not C\THC S.  A small effect may 
also be present in HC$_3$N in Taurus \citep{TakMas+98} and
\cite{MarAla+10} recently reported N(CCH)/N(\THC CH) 
$> 110 ~(3\sigma)$ in M 82.

The gas-phase carbon isotopic chemistry is the subject of this work.
Motivated by the apparent contradictions among the model 
predictions for a strong depletion of \THC , the absence of any 
obvious effect in subsequent measurements like those reported by 
\cite{MilSav+05} for CN and \THC N, and  the recent invocation of 
a general \THC\ depletion to explain anomalies seen in the nearby 
Taurus Cloud \citep{SakIke+07,SakSar+10}, we observed isotopologues
of species containing a single carbon (CS, HNC and \HH CS) toward 
three positions in Taurus, as described in Sect. 2.  Section 3 
summarizes the observational situation including our new measurements.  
Having found no evidence that \THC\ is depleted in the species we 
observed, and probably in the general pool of carbon in the gas, we 
discuss  the overall carbon chemistry in Sect. 4, inquiring how 
\THC\ depletion can be {\it avoided} and the implications for the 
gas phase chemistry if it does {\it not} actually occur.

\section{Observations and conventions}

The new observations reported here were taken at the ARO 12m telescope in
2011 March and April.  We observed HNC, CS, \HH CS and their 
singly-substituted \THC, \FTN\ and  \TFS\ isotopologues at three
positions in Taurus: the TMC1 cyanopolyyne peak 
($\alpha=4^{\rm h}41^{\rm m}42.88^{\rm s}$,
$\delta$=25\degr 41\arcmin 27\arcsec), L1527 
($\alpha=4^{\rm h}39^{\rm m}53.89^{\rm s}$,
$\delta$=26\degr 03\arcmin 11\arcsec) and the 
\ammon peak ($\alpha=4^{\rm h}41^{\rm m}23^{\rm s}$,
$\delta$=25\degr 48\arcmin 13.3\arcsec; all coordinates are J2000). All the
data were taken by position-switching to an off-position 8.5\arcmin\ NE
of the cyanopolyyne peak, using the dual-polarization mode of the autocorrelator
configured for 24.4 kHz channel spacing (0.073 \kms\ at 100 GHz) and 48.8
kHz spectral resolution.  Typical system temperatures were 180 - 240 K
depending on line frequency and source elevation.  All velocities are 
given with respect to the kinematic local standard of rest and the line 
temperatures are given in terms of the native scale at the ARO 12m,
\Tstar $ \approx 0.85$ T$_{\rm mb}$.  Line frequencies were taken 
from the online splatalog (http://www.splatalogue.net) and are given in Table 1
representing the new results.

In this work we do not call out the most-abundant isotope explicitly:
C implies $^{12}$C.  The integrated intensity of spectral lines for species
X, in units of K \kms, is denoted by W(X).

\section{Observational results}

\begin{figure*}
  \psfig{figure=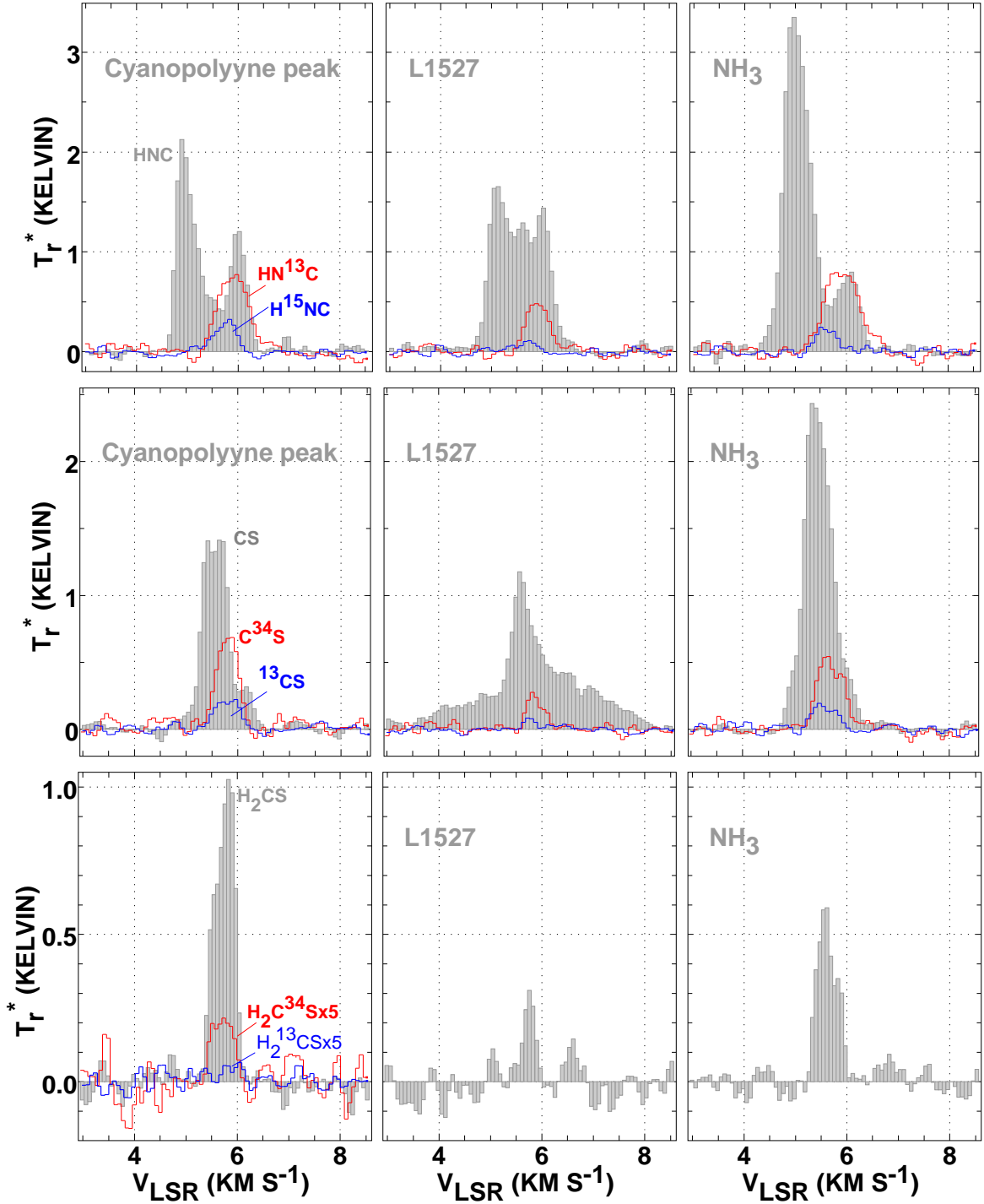,height=19cm}
  \caption[]{Observed line profiles toward three positions in TMC1} 
\end{figure*}

\subsection{New results}

Spectra are shown in Fig. 1 and numerical results are summarized in
Table 1 giving profile integrals and Table 2 giving ratios of line
profile integrals and implied carbon isotope ratios, assuming
values for N/$^{15}$N or S/$^{34}$S as indicated.

The main isotopologues of HNC and CS are optically thick and self-absorbed by 
weakly-excited molecules within the foreground envelopes; 
it is not possible to measure the C/\THC\ ratios in these species 
directly and we must assume values for the isotopologic ratios in 
sulfur and nitrogen.  For nitrogen the interstellar  N/\FTN\ ratio
is now seen to be only slightly larger than the terrestrial ratio of 
270. \cite{LisWoo+10} found N/\FTN\ = 334$\pm$50 ($3\sigma$) toward Barnard 1
and Adande and Ziurys (in press) find N/\FTN = 290 $\pm 40$ at the
Solar Circle from a large-scale galactic survey of CN emission.  
In local diffuse clouds \cite{LucLis98} found N/\FTN =240$\pm25$.
The range of values implied for HNC/HN\THC\ in Table 2 corresponds to 
HNC/H\FTN C = 250-330 and the statistical uncertainties are quoted at
each end of the range.

No systematic deviations from the Solar System value of S/\TFS = 22.7 have 
been seen in specific isotope studies of the local ISM  
\citep{Wan80,Wil99,TerCer+10} and the implied values of CS/\THC S quoted 
in Table 2 correspond to this value and the statistical 
uncertainties of the current measurements.  Our measurements of \HH CS
isotopologues toward TMC1 are in accord with this.  Some uncertainty accrues from 
the unmeasured optical depth of the rarer isotopologic lines as discussed 
in the following subsection.

The result for the isotopologic carbon ratio in \HH CS is of marginal 
significance and will not be discussed further but it seems clear from Table 
2 that there is no strong depletion of \THC\ in HNC or CS.

\subsection{Correction for finite optical depth}
\subsubsection{CS}

The presence of a finite optical depth in the C\TFS\ line would cause 
compression of the W(C\TFS)/W(\THC S) intensity ratio, resulting in an 
underestimate of the N(C\TFS)/N(\THC S) and  N(CS)/N(\THC S) column density 
ratios.
For TMC1, \cite{PraDic+97} derived N(C\TFS) = $2.0\times10^{12}\pcc$,
consistent with an estimate N(C\TFS) = N(CS)/22.7 = $5\times 10^{13}\pcc/22.7 
= 2.2\times 10^{12}\pcc$ from the work of \cite{OhiIrv+92}. 
For N(C\TFS) = $2.0\times10^{12}\pcc$ and peak line brightness of 0.72 K - 0.85 K,
depending on beam efficiency, we calculate that the central optical depth
of the C\TFS\ J=2-1 line toward TMC1 must be below 0.15 and the excitation
temperature in the lower part of the rotation ladder must be fairly high, 
above about 8 K for J=2-1, based on modelling of the excitation.
In this case the \THC S line is quite optically thin and the observed 
isotopologic intensity ratio is at most 5\% below the intrinsic abundance 
ratio.  Accordingly, the entry corrected for optical depth in Table 2 is
the uncorrected value increased by 5\%.

Estimates of the optical depth correction toward the other pointing positions 
do not seem possible.  Toward the NH$_3$ peak \cite{PraDic+97} derived a higher 
column density N(C\TFS) = $4.3\times10^{12}\pcc$ while the observed line in this 
work is only slightly broader and noticeably less bright.  Excitation 
solutions for such high column density in the face of relatively weak emission 
indicate high optical depth at lower excitation temperatures (5 K) that do not 
seem consistent with the physical conditions toward the \ammon peak, that are 
not less dense or colder than toward TMC1.  \cite{PraDic+97} cited problems
with their modelling of the CS lines, such that they were compelled to use
a very small isotopologic abundance ratio CS/C\TFS\ = 12-14.  Although
they cited a reference to the isotopic composition of cosmic rays in defense 
of this assumption, it is without precedent in molecular gas and 
would lead to improbably small values of the CS/\THC S ratio if used in this
work.  We conclude that  we cannot make a reliable correction for optical depth
correction except toward TMC1 where all indications are that the opacities
in the rarer CS isotopologues are small.

\subsubsection{HNC}

For HNC, the \THC-bearing isotopologue is the more abundant of the two variants 
used to derive the C/\THC\ ratio, so the effect of optical depth is opposite to
that in the case of CS: the derived HNC/HN\THC\ ratio decreases with increasing optical
depth in the rare isotopologues.  Indeed, the optical depths encountered in HNC are 
higher than for  CS and the correction for finite optical depth is rather larger.
Unfortunately we did not resolve the hyperfine structure of HN\THC and cannot
duplicate the excitation temperature/opacity derivation of \cite{PadWal+11} who 
observed at other positions in Taurus having higher optical depth than occurs
toward TMC1.

\cite{PraDic+97} gave N(HN\THC) = $4.2\times 10^{12}\pcc$ toward TMC1 while a 
slightly smaller upper  limit N(HN\THC) 
$< 2\times 10^{14}~\pcc/60 < 3.3\times 10^{12}\pcc$ can be inferred from the 
tabulation of \cite{OhiIrv+92}.  For the comparatively weak emission lines 
we observed, these column densities imply J=1-0 excitation temperatures around 
4 K and optical depths of order 2 in HN\THC, so that even such a rare isotopolgue
would be somewhat self-absorbed.  With optical depths this large, simply requiring 
that HNC/HN\THC\ be above 60 limits the possible range of assumed values for 
N(HNC)/N(H\FTN C) to be nearer the large end of the likely range.  
However, the small excitation temperatures implied by our observed brightnesses 
and the higher of the previously measured column densities  are not consistent 
with the densities that are inferred for the Taurus cores, which imply higher 
excitation temperatures, lower optical depths and smaller column 
densities. 

We conclude that it is not possible to derive an entirely accurate carbon isotopologue
ratio in HNC based on correcting the observed intensity ratios for optical depth
but the errors do go in a known direction.   Table 2 shows a range
of  carbon isotope ratios derived from HNC for an optical depth 0.6 in the 
HN\THC\ line toward TMC1, corresponding to an excitation solution with N(HN\THC) = 
$2\times 10^{12}\pcc$ that is consistent with the work of \cite{OhiIrv+92}.  This 
could still understimate the optical depths and so overestimate
the  C/\THC\ ratio. 

\subsection{Previously-reported \THC\ anomalies}

These are summarized in Table 3 where all the results are cast in terms 
of actual or implied C/\THC\ ratios, using the terrestrial sulfur 
isotopic ratio as appropriate.  There are two effects that must be considered: 
\THC\ is strongly lacking in some isotopologues but to different degrees
depending upon placement.  Thus both \THC CH and C\THC H are 
underabundant but moreso in the former (a factor 1.6) and in CCS 
only the more tightly-bound isotopolgue\THC CS exhibits the effect; 
in  each case, the isotopolgue with
an inboard \THC\ is more abundant.  In C$_3$S only \THC CCS
was observed, but is heavily deficient.  In terms of chemical models these 
two considerations are
separately reflected in, on the one hand, the overall C/\THC\ ratio in
the gas, and, on the other, the detailed path to formation (and perhaps
in situ fractionation after formation) of the individual species studied.

\cite{SakIke+07} and \cite{SakSar+10} discussed in exquisite detail the 
formation routes to CCS and CCH, concluding that the unequal depletions
of the  \THC-bearing isotopologues were the result of formation reactions
in which the carbon atoms were not equivalent 
(for CCH, C + C\HH\ $\rightarrow$ CCH + H and  not  C$_2$H$_3$\p\ + e 
$\rightarrow$ CCH + \HH), rather than alterations that might have
occurred after formation 
(for instance \THC CH + H $\rightarrow$ C\THC H + H + 8.1 K 
or \THC CS + S $\rightarrow$ C\THC S + S + 15 K).  They also
concluded that they had confirmed observationally the phenomena
modelled by \cite{LanGra+84} whereby the relatively famous
fractionation reaction involving CO and C\p\ induces a very 
general vanishing of \THC\ from the gas at large as discussed 
in the Introduction here.

\section{Carbon isotopic and isotopologic ratios in the Taurus cores}

The C/\THC\ ratios inferred from our work are consistent with other measurements 
of the carbon isotopic ratio in the nearby ISM that find C/\THC\ = 60-70 and 
show no sign of the \THC\ depletion that is predicted for the gas phase
chemistry.  The CS isotopologues we observed would not serve as the basis for substantially
depleted CCS but are instead close to the CCS/C\THC S ratio $54\pm2$
discussed by \cite{SakIke+07}, suggesting that CCS could indeed
form directly through a reaction like CH + CS $\rightarrow $ CCS + H as 
hypothesized by \cite{SakIke+07}. 

In fact, \cite{FurAik+11} recently reached something of the same conclusion
through chemical modelling, without the new observational result for CS.  
They considered a time-dependent model for TMC1 with all the carbon initially in 
the form of C\p\ (see \cite{Lis09}), in which the abundance of CO is relatively 
small even when it  begins to deplete out of the gas onto grains;  
the CO column in their model is also rather small compared with N(CO) = 
500 N(\coei) =$1.8\times 10^{18}\pcc$ from \cite{PraDic+97}.  Because of the 
assumed initial conditions a very large proportion of the gas-phase carbon 
remains outside CO, and, for reasons that are discussed in Sect. 5, \THC\ 
depletion in the gas as a whole is generally modest. C/\THC\ is 
typically of order 110 after the earliest times.  In this case the 
hypothesis of a rearrangement effected by reaction of CCS and H yielded 
CCS/\THC CS = 230, CCS/C\THC S = 54, as observed.  That is, the 
rare isotopologues both form with C/\THC\ = 110 but the hypothesized reaction 
C\THC S + H $\rightarrow$ \THC CS + 17.4 K is strongly energetically favored 
at 10 K driving CCS/\THC CS up  while CCS/C\THC S falls to half the C/\THC\ 
ratio in the gas (110/2).  Our results suggest that \THC CS and C\THC S 
actually form with CCS/\THC CS and CCS/C\THC S $\simeq 70$.

\cite{FurAik+11} also showed that the known rearrangement reaction for CCH with
ambient H (that has a smaller exothermicity 8 K) could yield unequal isotopologic 
abundances to the degree observed (a factor 250/170) but the relatively small amount
of \THC\ depletion in their model did not come close to reproducing the
very large individual CCH/\THC CH or CCH/C\THC H ratios that are observed 
(170 and 250; Table 3). Even so, their model generally predicts larger values for
N(CS)/N(\THC S) than we observe.    It is not obvious that this scheme would function 
as well in a gas that was even more weakly fractionated, if it caused very small values 
for the CCS/\THC CS rato.

The smaller isotopologic discrepancies that are observed in HC$_3$N can likely 
be  achieved by relatively weakly-endothermic rearrangement reactions with hydrogen in
an unfractionated gas, consistent with the lack of fractionation we infer for HNC.
In summary, the observational situation is complex and somewhat mysterious.  CS, HNC and
HC$_3$N suggest that there is little fractionation in the carbon pool at large while
the seeming absence of fractionation in CCS/C\THC S may be accidental and the anomalies
in CCH remain unexplained.  Disparities
among the carbon isotopologues in individual species with more than one carbon atom seem best
explained by rearrangement reactions with atomic hydrogen but the very large isotopologic
ratios seen in CCH remain beyond the reach of current explanations and the disparity that
is observed in CCS may be inconsistent with rearrangement in a gas that is too-lightly
fractionated.

\section{Where is the CR ionization-driven fractionation?}

The possibility of a strong overall carbon 
fractionation in species other than CO has been recognized ever since 
the existence of the C\p\ + CO fractionation reaction was first remarked. 
Nonetheless, it was not observed in surveys that sought to determine the 
carbon isotope ratio but found the same values in CO and CS and CN etc, 
even though the latter species do not form from CO and could
have shown strong fractionation effects even in relatively warm gas.  
Even now, despite the carbon isotopologic anomalies that are seen at
very low temperatures (10 K) in Taurus in some polyatomics
bearing more than one carbon, the gas there can not obviously be said 
to be fractionated as a whole.  Here we ask why this is so.

\subsection{The role of He\p}

The prediction of strong overall fractionation arises in models of the gas-phase 
chemistry of dense dark oxygen-rich material (O/C $>$ 1) where CO is by far the 
largest repository of gas-phase carbon atoms \citep{LanGra+84}. 
A few molecular species form
directly from CO -- \hcop\ in the gas, \HH CO and CH$_3$OH from 
successive hydrogenation of CO on grains -- and will share its 
C/\THC\ ratio.  Other carbon-bearing trace molecules  
in the gas result from the relatively small amount of carbon that 
is liberated when He\p, ionized by cosmic rays, slices CO apart 
into C\p\ and O via the reaction He\p\ + CO $\rightarrow$  
C\p\ + O + He; the  reaction rate constant is $1.6\times10^{-9}\ccc~\ps$ 
(see Table 4 for important reactions and rate constants).  The 
only reaction competing effectively with this one for the attentions 
of He\p\ is the charge-exchange ionization of \HH,  
He\p\ + \HH\ $\rightarrow$ He + \HH\p, whose currently-accepted
reaction rate constant is $7.2\times10^{-15}\ccc~\ps$ (Table 4).
\cite{LanGra+84} included instead the reaction 
He\p\ + \HH\ $\rightarrow$  H\p\ + H + He with the much larger rate 
constant 1.5$\times10^{-13}\ccc~\ps$, which significantly 
dampened the effects of fractionation compared to what would be 
derived in more recent chemical models: that reaction is now considered 
to be negligibly slow in cold gas, see Table 4.

The ratio of reaction rates and abundances determines whether CO or \HH\ 
controls the abundance of He\p\ but for X(CO) $> 4.5\times10^{-6}$, when 
CO dominates using currently-accepted reaction rate constants, every 
cosmic-ray ionization of an He atom results in the production of a C\p\ ion.  
Because He is so much more abundant than carbon (i.e. CO), the volume 
rate of production of C\p\ ions and the rate at which carbon is liberated 
from CO is approximately 1000 times higher than the rate at which 
cosmic rays interact with CO directly; this factor is independent
of the cosmic ray ionization rate n(He\p) and independent of X(CO) 
when the reaction with CO is the main destroyer of He\p. 

It is important to stress just how rapidly the chemical ionization of 
carbon occurs via He\p\ + CO,  because the only competing mechanism 
for liberating atomic carbon from CO is photodissociation.  In quiescent, 
shielded, cosmic-ray heated gas, cosmic-ray induced photodissociation
typically occurs at rates that are a much smaller multiple of the cosmic ray 
ionization rate, and is some 50 times slower than He\p + CO.  
The point is that in these models it is difficult to contrive to 
liberate carbon from CO except, initially, as C\p, leading to 
fractionation.

\subsection{The role of the CO + C\p\ fractionation reaction}

Just as He\p\ reacts most commonly with CO, so does C\p, although in this 
case there is a wider range of competitive reactions (see Table 4).  
But to the extent that C\p\ interacts with CO, it preferentially deposits 
\THC\p\ back in CO.  Meanwhile, the abundance of C\p\ is so small relative 
to that of CO (typically less than 0.1\%) that any effect on the isotopologic 
abundance ratio in CO is imperceptible.  The final result is that when free 
atomic carbon is primarily liberated and maintained in the form of C\p, 
the  pool of gas-phase carbon outside CO inevitably becomes depleted in \THC\p,
hence in \THC,  as long as CO remains sufficiently abundant in the gas 
to dominate the neutralization of He\p.  CO and the free carbon pool
coexist but with two distinct C/\THC\ ratios that are separately  passed 
on to their descendant molecules.


\subsection{So indeed, why is a more general \THC\ depletion not seen in Taurus?}


To blunt the effect of C\p\ fractionation it is necessary to interfere
with the C\p-CO interaction and there are various choices for doing so,
for instance; i) a low CO abundance, such as might occur at later 
times after depletion onto dust (see also the early-time model of 
\cite{FurAik+11} in which CO never attains its full abundance); 
ii) very large abundances of species like OH and O$_2$ that react 
rapidly with C\p; iii) taking the high-metals case where sulfur and 
silicon are not assumed to be strongly depleted from the gas so that they
dominate the neutralization of C\p\ through charge exchange;  
iv) the presence of PAH \citep{WakHer08,LepDal+88} to neutralize
C\p\ more rapidly, although these are not always assumed to survive in dense 
gas.   Nonetheless, fractionation at the level of a factor two or so 
persists in the C\p\ until CO and C\p\ are all but gone from the gas.

Preventing fractionation of the pool of molecules outside the small
family that forms from CO really requires that the reservoir 
of carbon outside CO must reside mostly in the form of neutral atomic 
carbon with perhaps some small admixture of C\p.  Given that C\z\
typically reacts with neutral molecules at least 20 times more slowly
than does C\p, the proportion of C\z\ to C\p\ should be very large and 
the real question is how this might come about.  
One possibility that is
capable of maintaining very large free neutral atomic carbon fractions is 
the so-called high ionization phase of bistable solutions of the chemical 
equilibrium equations, whose relevance has recently been discussed in terms 
remarkably similar to the considerations here for controlling the C\p\ abundance
\citep{ChaMar03,WakHer+06,BogSte06}.  Extension of the bistability discussion
to include fractionation effects is clearly of great interest.

It is  generally {\it not} possible to create a sufficient pool of
neutral carbon using just the in situ flux of CO-dissociating photons generated 
by the cosmic ray-induced electronic excitation of \HH\ \citep{PraTar83,SteDal+87} 
because these photons dissociate CO at rates that are a relatively small multiple 
of the cosmic ray ionization rate: \cite{LanGra+84} assumed a multiple of 10.  
As such, cosmic-ray induced photons liberate neutral atomic carbon much 
more slowly than CO is sliced into C\p\ and O by He\p. 
Because the naturally-arising local flux of CO photodissociating photons 
inside dark gas is so weak, some attempts to explain the existence of even a 
relatively small amount of C\z\ in dense gas invoke special mechanisms 
such as inversion of the normal [O]/[C] $>$ 1 ratio \citep{LanGra+84}, 
penetration of ambient uv light into a porous, heavily clumped medium
and/or recent cloud formation \citep{PhiHug81},
turbulent diffusion that cyclically exposes heavily-shielded gas to the ambient 
uv radiation field near the cloud surface  \citep{BoldeJ82,WilLan+02,XieAll+95} 
and the presence of the
so-called high-ionization phase of bistable chemical reaction schemes
\citep{FloLeB+94}.

The inference of a large pool of free atomic carbon in dense dark gas is really
not an observational problem per se because substantial columns of C\z\ are 
typically found toward and around dark clouds \citep{FreKee+89} including 
toward TMC1 where N(C\z)/N(CO) $\simeq 0.1 $ \citep{MaeIke+99}.  This is very 
large compared to the values X(C\p)/X(CO) 
$\sim 3\times10^{-8}/8\times10^{-5} \sim 4\times10^{-4}$ that arise from 
the default parameters in our toy model (Appendix B).  However, the 
observational situation is complicated by superposition of different
density regimes along the line of sight and it is hard to assess how 
much of the observed atomic carbon actually exists inside the darkest gas.  
This is especially true given the recent revelation of the preponderance
of low-\AV\ material in Taurus and other dark cloud complexes 
\citep{PinGol+10,Cam99}.  
Extrapolation of the results of \cite{BenLeu+03} to much higher N(CO) 
suggests that perhaps 1-2\% of the carbon budget \citep{GerFos+03} 
could exist as C\z\ at N(CO) $\approx 10^{18}\pcc$.  This would be 
about 50 times more carbon than exists in C\p.

\subsection{Relevance to other environments}

The possibility of chemical fractionation in cold cores raises important 
questions for understanding carbon isotopic abundances in planetary and 
proto-planetary systems including the Solar System and proto-Solar nebula.  
For example, is it correct to ascribe the difference between the Solar ratio 
C/\THC\ = 89 and that in the nearby ISM (60-70) entirely to chemical enrichment 
since the birth of the Solar System or might fractionation be responsible
\citep{SmiPon+11} ?  
Why are there differences between the carbon isotope ratios measured in 
various comets and between those measured in comets and that in the Sun
\citep{CroBiv+09,MumCha11}.  Other questions arise in matters involving other 
elements, for instance the difference between the N/\FTN\ ratio in the Earth 
(270) and Sun (420) and the wide disparities in D/H measurements between that 
in the earth's oceans ($1.6 \times 10^{-4}$), the more nearly cosmological 
ratios seen in the outer planets ($3-5\times 10^{-5}$) and the much higher 
than telluric D/H ratios seen in most comets \citep{CroBiv+09,MumCha11},
although, apparently, not all \citep{HarLis+11}.

Many of the effects discussed here occur prominently in recent models of 
fractionation chemistry in protoplanetary disks \citep{WooWil09}, which 
illustrate the complexity of relating isotope ratios in the proto-planetary 
nebula or planetary disk (when formed) to that in the ambient natal material.  
Fractionation varies with disk radius and disk height and evolves with time 
under the combined influence of the proto-stellar and interstellar radiation 
fields.   \cite{WooWil09} conclude that Solar System cometary material has 
been reprocessed, raising the question of whether any memory of conditions 
in natal molecular material persists into fully-formed planetary systems.

\section{Summary}

Since the initial recognition of the carbon isotope fractionation reaction 
a conflict has existed between the very general prediction of a strong
\THC\ depletion in molecules other than CO and the general absence of 
observable effects in surveys of the C/\THC\ isotopic abundance ratio 
deduced form such common species as CS or CN.  \THC\ depletion is predicted 
to occur when carbon is liberated from CO by the reaction 
He\p + CO $\rightarrow$ C\p\ + O + He and remains in the gas as C\p\ 
to interact and fractionate with CO.  Eventually, CO depletion onto 
grains will blunt the effect at late times but in the meanwhile quite 
large variations in the abundances of \THC-bearing molecules are 
predicted for all species that do not form directly from CO.  The point 
is that CO remains very nearly unfractionated as long as it is the main 
carbon reservoir in the gas and X(C\p)/X(CO) is very small, so that species 
like \hcop, \HH CO and CH$_3$OH will also be unfractionated. 
But the relatively small pool of carbon that exists outside of
CO to form other molecules than those in the CO family becomes strongly depleted 
in \THC\p\ and molecules that form from it (most carbon-bearing species)  will also 
show strong \THC\ depletion.

However, strong anomalies in CCH and in some isotopologues of CCS and 
CCCS have recently been observed in cold dark gas in TMC1 and other
cores in Taurus (less strongly in HC$_3$N there). It was suggested that the
predicted fractionation effects had at long last actually been  seen, 
albeit in only some very cold gas.
We showed here that \THC\ fractionation does not occur in two species
having a single carbon (CS and HNC) seen toward TMC1 in Taurus; we also 
observed these species toward other positions, finding certainly no
clear evidence of fractionation, but derivation of an accurate 
isotopic abundance was frustrated by the difficulty of correcting for
finite optical depth even in very rare isotopologues.  In any case, our
observations toward TMC1 make it unlikely that any general depletion
of \THC\ exists in the gas at large outside CO and we discussed the 
implications of this inference in the context of the chemistry of
an optically shielded dense gas with a normal ratio [O]/[C] $>$ 1.

In general, preventing an overall \THC\ fractionation requires that
the pool of gas-phase carbon outside CO resides in the gas mostly as C\z\
rather than C\p. This could happen if the [O]/[C] ratio is 
less than unity or if some mechanism is invoked to liberate 
neutral carbon from CO through photodissociation deep inside dark gas
but the most interesting possibility is that
the TMC1 gas is in the so-called high ionization state of a bistable
chemical network.  In any case, we are left with the fact that the 
overall carbon pool outside CO is apparently not depleted in \THC\ even if 
there is no way to prevent \TWC\p/\THC\p $>>$ 60.  For the chemistry 
this has interesting
consequences; reactions involving C\p\ see a pool of carbon
depleted in \THC\ while those with C\z\ do not, and the main pool
of carbon is in larger amounts of C\z\  that react somewhat more 
slowly, with a normal \TWC/\THC\ ratio.

In the future it may be possible to ascertain the overall composition of
the carbon pool deep inside TMC1 using high spatial resolution observations 
of sub-mm lines of C I at ALMA, and perhaps even to measure the C/\THC\ 
ratio using the 492.164 GHz line of \THC\ that is displaced from the
main isotope.
In the meantime it is important to assess just which chemical species are
subject to isotopologic abundance anomalies - apparently, a wide variety
of tri- and polyatomics hosting more than one carbon atom  -- 
and under what conditions they arise, and to avoid using strongly affected 
molecles to derive isotopic abundance ratios.

\acknowledgments

   The Arizona Radio Observatory is operated by Steward Observatory, 
  University of Arizona, with partial support through the NSF ATI program.  
  The National Radio Astronomy Observatory is operated by Associated
  Universites, Inc. under a cooperative agreement with the US National
  Science Foundation.  HSL wishes to acknowledge the hospitality of
  Gasthaus Leykauf in Rottach-Egern during the completion of the initial
  draft of this work and that of Bel Esperance in Geneva during the
  next.  We thank the referee for a close reading of the manuscript
  and a variety of interesting suggestions for further discussion.



{\it Facilities:} \facility{ARO}



\appendix

\section{Appendix A. The fractionation reaction}

\cite{SmiAda80} did not give analytic forms for their measurements
of the  temperature dependence of the carbon fractionation reaction
first cited by \cite{WatAni+76}.  \cite{Lis07} provided the expressions

$$ \kf = 7.64 \times 10^{-9} ~\TK^{-0.55} \ccc~\ps ~~
(\TK = 80-500 {\rm K}) $$

$$ \kf = {1.39 \times 10^{-9} ~\TK^{-0.05} \ccc~\ps  \over {1+\exp{(-34.8/\TK)}}}
  ~~( \TK = 10 - 80 {\rm K}) $$

$$ \kr = \kf \exp{(-34.8/\TK)} $$

for the forward (exothermic) and reverse reaction rates $\kf$ and $\kr$.

\section{Appendix B. A toy model of \THC\ depletion}

\begin{figure}
\psfig{figure=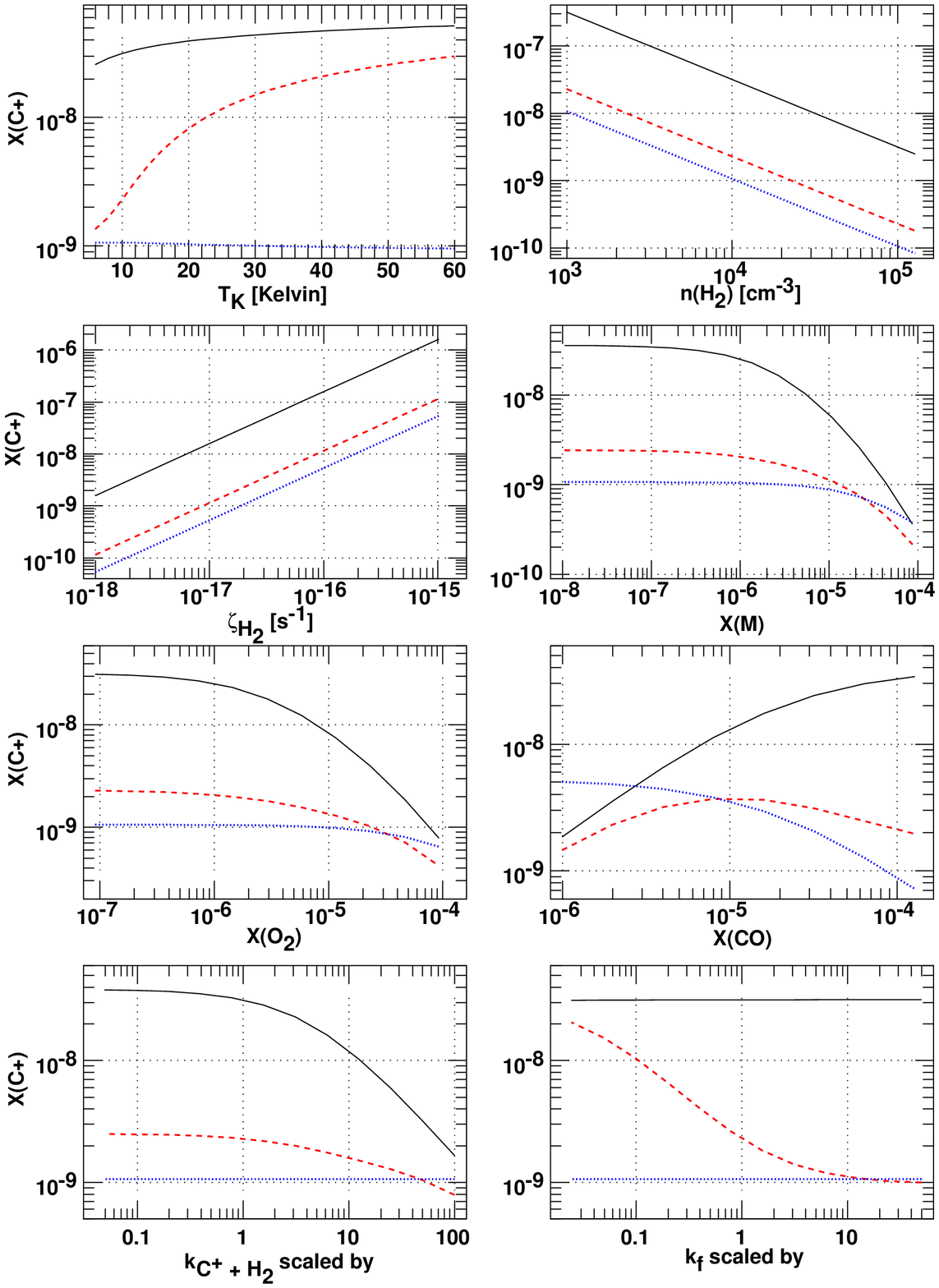,height=20.5cm}
\caption[]{Toy model of the \THC\p , \TWC\p\ chemistry.  In each panel
the solid line is X(C\p), the (red) dashed line is 
[\TWC]/[\THC]$\times$ X(\coth)/X(\cotw) and the (blue) dotted line
is X(He\p) (only shown in some panels).  Standard values are 
\TK\ =  10 K, n(\HH) = $10^4\pccc$, $\zeta = 2\times 10^{-17}$ per \HH,
 X(CO)= $8\times10^{-5}$, X(M) = $3\times 10^{-7}$, 
X(O$_2$) = $3\times 10^{-7}$ and \TWC/\THC\ = 60.  The panels of this 
figure show the effects of varying individual parameters.  In the bottom
two panels two important rates are scaled by the amounts shown on the
horizonal axis.}
\end{figure}

We constructed a formal model of the chemistry in which 
C\p\ is produced by the action of He\p\ on a fixed abundance of CO 
(X(CO)= n(CO)/n(\HH) is a parameter) and solved for the densities 
n(\TWC\p), n(\THC\p) and  n(He\p).  In addition to the fractionation 
reaction with CO, C\p\ is destroyed by interaction with \HH, hydrides 
(CH, NH, OH), O$_2$, thermal electrons, and charge exchange with a single 
low-ionization metal species denoted ``M'' and given the atomic
properties of silicon.   For the reactions in Table 4 
we used the dipole-enhanced rates given on the UFDA06 database website 
{\it http://www.udfa.net/} \citep{WooAgu+07}.

The default values of the reactant abundances are 
given in Table 4 and quoted in Figure 2, i.e. n(\HH) = $10^{4}\pccc$, 
\TK = 10 K, \zetaMolH = $2\times10^{-17}\ps$.  For the default electron 
abundance $2.4\times10^{-7}$ we used the expression in \cite{OppDal74}; 
alternatively see \cite{McK89}. The default carbon isotopic ratio is 
\RC = [$^{12}$C]/[$^{13}$C] = 60.   The entry for 'M' represents all heavy atoms with 
ionization potentials less than 10ev or so and the default 
is for strong depletion -- the typical ``low metals'' case.  Nitrogen
was assumed to exist in the form of N$_2$ with a Solar [N]/[C] ratio
and carbon depletion in the model corresponds to the default value X(CO) =
$8\times10^{-5}$.

For He\p
$$dn({\rm He}\p)/dt = \zeta_{\rm He} n({\rm He}) 
 - n({\rm He}\p) n(\HH) \sum_j{k_j  X_j} $$

where the fractional abundances of the reactants j are X$_j$ and their 
reaction rate constants with He\p\ are k$_j$ (listed in Table 4).  
For He we take the local galactic disk abundance [He]/[H] = 0.088 \citep{Bal06}.
The cosmic-ray ionization rate of He is $\zeta_{\rm He}$  = 1.08 \zetaMolH/2 
and \zetaMolH = $2\times10^{-17}~\ps$.  In general, direct recombination
of atomic ions with ambient electrons is utterly insignificant in this context.

For \TWC\p\ and \THC\p\
\begin{eqnarray*} dn(\TWC\p)/dt = n(\cotw) [n(\THC\p) \kf + n({\rm He}\p)\kHeCO] \\
-n(\TWC\p)[n(\coth)\kr +  n(\HH) \sum_j{\kjCp  X_j}] 
\end{eqnarray*}
\begin{eqnarray*} dn(\THC\p)/dt = n(\coth) [n(\TWC\p) \kr + n({\rm He}\p)\kHeCO] \\
-n(\THC\p)[n(\cotw)\kf +  n(\HH) \sum_j{\kjCp  X_j}]
\end{eqnarray*}

In each panel of Fig. 2 the relative abundance 
X(\TWC\p) is shown by a solid (black) line and X(\THC\p) $\times$ \RC\ 
(the red dashed line) has been scaled up by the inherent isotopic abundance 
so that the gap between the solid black and dashed red lines shows
directly the extent to which \THC\ is depleted in C\p\ and in the gas
outside CO.  The effect is generally very large at 10 K, though still 
somewhat below exp(35/\TK).  The lower right panel shows the effect of 
artificially increasing the strength of the fractionation reaction, 
illustrating that depletion of \THC \p\  might be a factor two stronger if 
competing reactions were less important. 

Although the important chemical effects ahould be incorporated in the 
default model, its \THC\ depletion (a factor 15) is much stronger than is seen 
even in CCH (at most a factor 4) and substantial \THC\ depletion 
persists at relatively high temperature (the upper left panel).  Of course
none of this is observed.  Lowering
the C\p/\THC \p\ ratio can be accomplished ad hoc by assuming an 
undepleted metal abundance, by putting all the spare oxygen in a species 
like O$_2$ or OH that reacts rapidly with C\p, or by increasing the
rate at which C\p\ recombines with \HH\ (lower left).  Perhaps the least
ad hoc modification is the case of small X(CO) because CO will eventually 
deplete from the gas phase.  This modificatin is basically the effect suggested
by \cite{Wat77} and modelled by \cite{Lis78}.

\section{Appendix C. Is carbon fractionation reversible after formation?}

In the fractionation chemistry the ambient \THC\p\ is depleted by a factor somewhat smaller 
than exp(35 K/\TK) corresponding to the binding energy difference in CO.  
In principle, if another molecule's chemistry was dominated by a fractionation 
reaction like that of C\p\ and CO, with an energy defect comparable to 35 K 
and an abundance much less than that of C\p\ (which itself is not large) so that
the gas contains sufficient amounts of \THC, that molecule's 
complement of \THC\ could be restored to a degree approaching the 
abundance ratio inherent in the gas, \RC.  
At present it is not possible to demonstrate that such in situ 
fractionation after formation occurs for any species beside CO: species 
either react chemically with C\p\ to form other species or they 
react too frequently with other things beside C\p, with the added 
complication that the rate constants of the required fractionation 
reactions are still unknown even 35 years since the importance of the CO
fractionation reaction was established.  CS is in fact one of the 
better chemical candidates, 
with an energy difference between CS and \THC S of 26.3 K, but
the fractionation reaction with C\p\ has never been measured.


\bibliographystyle{apj}



\begin{table}
\caption[]{Species, line frequencies and integrated line temperatures$^a$}
{
\begin{tabular}{lcccc}
\tableline
Species & Frequency & Cyano Pk& L1527 & \ammon Pk \\
\tableline
       & MHz & K-\kms & K-\kms & K-\kms \\
\tableline
HNC    & 90663.56 & 1.597(0.020) & 1.850(0.040) & 2.550(0.020) \\
HN\THC & 87090.85$^b$ & 0.551(0.023) & 0.263(0.022) & 0.650(0.026) \\
H\FTN C & 88865.69 &0.120(0.016) & 0.055(0.015) & 0.086(0.016) \\
       &  & & \\
CS   & 97980.95 & 0.980(0.011) & 1.495(0.023) & 1.685(0.010) \\
C\TFS & 96412.95 &0.362(0.014)& 0.116(0.009) & 0.341(0.005) \\
\THC S & 92494.27 &0.121(0.008) & 0.045(0.007) & 0.100(0.008) \\
       &  & & \\
\HH CS  & 103040.28& 0.483(0.012) & 0.102(0.019) & 0.272(0.011) \\
\HH C\TFS & 101284.40 & 0.0218(0.0044) & & \\
\HH \THC S & 99077.84 & 0.0063(0.0016)& & \\
\tableline
\end{tabular}}
\\
$^a$ Table entries are $\int \Tstar dv$ quantities in parentheses are 1$\sigma$ uncertainty  \\
$^b$ For details of the HN\THC\ spectrum see \cite{vanMul+09} \\
\end{table}

\begin{table*}
\caption[]{Integrated temperature and implied isotopologic ratios}
{
\small
\begin{tabular}{lccc}
\tableline
Species & Cyano Pk & L1527 & \ammon Pk \\
\tableline
W(HN\THC)/W(H\FTN C) & 4.59(0.64) & 4.78(1.36) & 7.56(1.44) \\
N(HNC)/N(HN\THC )$^a$ & 54(8)..72(10) & 52(15)..69(20)& 33(6)..44(8) \\
N(HNC)/N(HN\THC )$^b$ & 43(6)..57(9) & & \\
       &  & & \\
W(\THC S)/W(C\TFS) &0.334(0.024) & 0.385(0.066) & 0.293 (0.028) \\
N(CS)/N(\THC S)$^c$  & 68(5) & 59(10) & 77(7) \\
N(CS)/N(\THC S)$^d$  & 71(5) & & \\
       &  & & \\
W(\HH CS)/W(\HH C\TFS) & 22.9(4.2) & & \\
       &  & & \\
W(\HH \THC S)/W(\HH C\TFS) & 0.289(0.094) & & \\
N(\HH CS)/N(\HH \THC S)$^e$ & 79(26) & & \\

\tableline
\end{tabular}}
\\
$^a$ for N(HNC)/N(H\FTN C) = 250..330 and $\tau = 0$ in the HN\THC\ line \\
$^b$ for N(HNC)/N(H\FTN C) = 250..330 and $\tau = 0.6$ in the HN\THC\ line. \\
$^c$ for N(CS)/N(C\TFS) = 22.7 and $\tau = 0$ in the C\TFS\ line \\
$^d$ for N(CS)/N(C\TFS) = 22.7 and $\tau = 0.15$ in the C\TFS\ line \\
$^e$ for N(\HH CS)/N(\HH C\TFS) = 22.7 and $\tau = 0$ in the \HH C\TFS\ line.
\end{table*}

\begin{table}
\caption[]{Other measured and inferred \TWC/\THC\ ratios$^a$}
{
\begin{tabular}{lccccc}
\hline
Species & Cyano Pk  &1521E & L1527 & \ammon & ref \\
\hline
CCH/\THC CH & $>250^b$ & & $>135^c$ & & 1 \\
CCH/C\THC H & $>170^b$ & & $>80^c$ & & 1 \\
  & & & & & \\
CC\TFS/\THC CS & 230(43) & $>$130 & & & 2 \\
CC\TFS/C\THC S & 54(2) & 51(4) & & & 2 \\
  & & & & & \\
C$_3$\TFS /\THC CCS  & $>191$ &  $>107$ & & & 2 \\
  & & & & & \\
HC$_3$N/H\THC CCN & 79(11) & & & & 3 \\
HC$_3$N/HC\THC CN & 75(10) & & & & 3 \\
HC$_3$N/HCC\THC N & 55(7) & & & 45(6) & 3 \\
\hline
\end{tabular}}
\\
$^a$ Lower limits are $3\sigma$, rms in parentheses are $1\sigma$ \\
$^b$ C\THC H/\THC CH = 1.6 $\pm0.4(3\sigma)$ \\
$^c$ C\THC H/\THC CH = 1.6 $\pm0.1(3\sigma)$ \\
References:
1) \cite{SakSar+10} \\ 
2)  \cite{SakIke+07} \\
3) \cite{TakMas+98} \\
\end{table}

\begin{table*}
\caption[]{reactants and their rate constants with He\p\ and C\p\ }
{
\begin{tabular}{llcc}
\hline
Reactant & Relative abundance & w/He\p     & w/C\p  \\
         &           & $\ccc~\ps$ & $\ccc~\ps $\\
\hline
\HH & 1 & \tim7.2,{-15}+\tim3.7,{-14} $\exp{(-35/\TK)}$
 & \tim4.0,{-16} $(300/\TK)^{0.20}$\\
CO & \tim8.0,{-5} & \tim1.6,{-9} & see A1\\
e  & \tim2.4,{-7} & & \tim7.2,{-12} $(300/\TK)^{0.83}$$^a$ \\
M & \tim3.4,{-7}  & \tim3.3,{-9} & \tim1.5,{-9} \\
CH & \tim1.0,{-7} & \tim1.6,{-9}\ft & \tim3.8,{-10}\ft \\
OH & \tim3.0,{-7} & \tim1.0,{-9}\ft &  \tim7.7,{-10}\ft \\
O$_2$& \tim1.0,{-7} & \tim1.0,{-9} & \tim1.0,{-9} \\
\HH O&\tim1.0,{-7} & \tim4.8,{-10}\ft & \tim2.7,{-9}\ft \\
N$_2$&\tim1.2,{-5} &\tim1.6,{-9} &\\
\hline
\end{tabular}}
\\
$^a$ \cite{WolTie+08}, includes dielectronic recombination
\end{table*}

\end{document}